# Towards the Secure Storage of Images on Multi-Cloud System

Dr. Grasha Jacob[1], Dr. A. Murugan[2]

[1]Associate Professor, Dept. of Computer Science,
Rani Anna Govt. College for Women, Tirunelveli, India
[2]Associate Professor, Dept. of Computer Science,
Dr. Ambedkar Govt Arts College, Chennai, India
grasharanjit@gmail.com

**Abstract-** *With the rapidly changing technological realm, there is an urgent need to provide and protect the confidentiality of confidential images when stored in a cloud environment. To overcome the security risks associated with single cloud, multiple clouds offered by unrelated cloud providers have to be used. This paper outlines an integrated encryption scheme for the secure storage of confidential images on multiple clouds based on DNA sequences.*

## 1. INTRODUCTION

Cloud computing has been defined as a model for enabling convenient, on-demand network access to a shared pool of configurable computing resources that can be rapidly provisioned and released with minimal management effort or service provider interaction. But there is a risk of widespread data loss or downtime due to internet connectivity problems or a localized component failure in a cloud computing environment. In order to reduce the service availability risk or loss of data, a multi cloud system where data is replicated into multiple clouds offered by different cloud providers are used. Once the assets are in the cloud, the clients lose control over them, as data in the cloud typically resides in a shared environment. With the growth of a large number of cloud service providers, there is a need to protect confidential images stored in the cloud from unauthorized users and also from the cloud service providers. This paper proposes an integrated encryption scheme based on DNA sequences and scrambling using zigzag pattern or Magic Square of doubly even order pattern to ensure double-fold security for the secure storage of confidential images on multi-cloud.

## 2. DEFINITIONS FOR THE PROPOSED SYSTEM

2.1. **Cryptosystem :** A cryptosystem is a five tuple (M, C, κ, ε, D) where

i. M is a finite set of plain-text (images).
ii. C is a finite set of possible cipher-text.
iii. κ is a finite set of possible keys.

iv. For each K ∈ κ, there is an encryption rule, $E_K \in \varepsilon$ and a corresponding decryption rule, $D_K \in D$.

Each $E_K : M \rightarrow C$ and $D_K : C \rightarrow M$ are functions such that $D_K(E_K(x)) = x$, for every plain-text $x \in M$.

**2.2. Synthesis:** Creation of DNA sequences for the image data is referred to as Synthesis. DNA sequences are made up of four bases – A, C, T and G. According to the DNA Digital Coding Technology [6], C denotes 00, T – 01, A - 10 and G – 11.

**2.3. Translation:** When the positions of sequences are translated, the sequences are interchanged.

$$P_1P_2P_3P_4 \leftrightarrow P_5P_6P_7P_8$$

**2.4. Substitution:** Each quadruple nucleotide sequence is substituted by the value returned by the DNA Sequence Crypt function.

$$V \leftarrow DNASequenceCryptfn(P_1P_2P_3P_4)$$

**2.5. Detect:** Detect searches for a quadruple nucleotide sequence of the image starting from a random position in the DNA sequence file and returns true if a match is found and false otherwise.

Boolean ← Detect($P_1P_2P_3P_4$)

**2.6. DNA Sequence Crypt function:** A DNA Sequence Crypt function is a one-to-many function d(x) that returns one of the many positions of the quadruple DNA sequence in the key DNA sequence file.

**2.7. Rev-Synthesis:** Rev-synthesis is the process of converting each sequence into its digital form.

**2.8. Magic Square:** A magic square of order n is a square matrix or array of $n^2$ numbers such that the sum of the elements of each row and column, as well as the main diagonals, is the same number, called the magic constant, σ(M). Figure 1 displays magic squares of order 8 arranged in eight different patterns.

**2.9. Zigzag ordering:** An n x n image arranged in a zigzag pattern is referred to as zigzag ordering. Figure 2 represents eight different ways of zigzag ordering.

## 3. PROPOSED SCHEME

In the proposed integrated encryption scheme, each pixel of the digital image is converted into its corresponding DNA coded value according to the DNA encoding method proposed for each pixel of a digital image by Qiang Zhang et al[2].

The integrated encryption scheme consists of two phases. Phase I deals with the secure storage and Phase II deals with the retrieval of the stored image. Figure 3 pictorially represents Phase I. In Phase I, the image to be encrypted is first synthesized - transformed into DNA image. An intermediate image is then obtained by substituting each quadruple DNA nucleotides sequence of the translated image by one of the many positions of the quadruple nucleotides sequence which is randomly obtained from the gene sequence file. The intermediate image is further scrambled according to magic square of doubly even order pattern or zigzag pattern to ensure double-fold security, and the resultant encrypted image which is of the same size as that of the original image is stored in the cloud. In the retrieval phase (Phase II), the authorized receiver upon receiving the encrypted image, re-scrambles it, converts it into DNA image by mapping the pointers from the encrypted image onto the key DNA sequence file and the decrypted image is then obtained by Rev-Synthesis. Phase II of the integrated encryption scheme is given in Figure 4.

Both the DNA Sequence file used as the key sequence and the scrambling pattern are selected at random and are embedded in the corner bits of the encrypted image. The authorized user agrees upon a look-up table for the scrambling pattern and gets the indices for both the scrambling pattern and the key DNA Sequence file from the encrypted image itself. Each time when the image is stored, a different scrambling pattern and a different key DNA Sequence file chosen at random are used to ensure security and uniqueness of encrypted images in multi cloud. In a multi-cloud environment, for the same image the encrypted images are different(though they look similar) in different clouds as the scrambling pattern and the key DNA sequence files are different.

## 4. RESULTS AND SECURITY ANALYSIS

To prove the strength and soundness of the integrated encryption scheme, experiments were performed on 200 different images of varied sizes( 64 x 64, 128 x 128, and 512 x 512) from different data sets(medical records, defence and personal images) using Matlab 2009a on DELL Inspiron ACPIx64 based notebook PC.

Security analysis is made to find the weakness of a cryptographic scheme and retrieving the encrypted image without the knowledge of the decryption key.

### 4.1 Correlation Coefficient

The Correlation Coefficient is determined using the formula:

$$r = \frac{n\sum xy - (\sum x)(\sum y)}{\sqrt{n(\sum x^2) - (\sum x)^2}\sqrt{n(\sum y^2) - (\sum y)^2}} \quad\quad (1)$$

where x and y are two adjacent pixels and n is the total number of pixels selected from the image for the calculation. Figure 5 (b), (d) and (f) expose that the correlation of the encrypted images is uniformly distributed and divulges no information to any unauthorized users. Table 1 elucidates that there is negligible correlation between the adjacent pixels of the encrypted images.

### 4.2 Histogram

Histograms express the statistical characteristics of an image. An encryption algorithm has good performance if the histograms of the original image and encrypted image are dissimilar. Figure 6(b) and (c) reveal that the histograms of the original and encrypted images are entirely different and an unauthorized user will find it

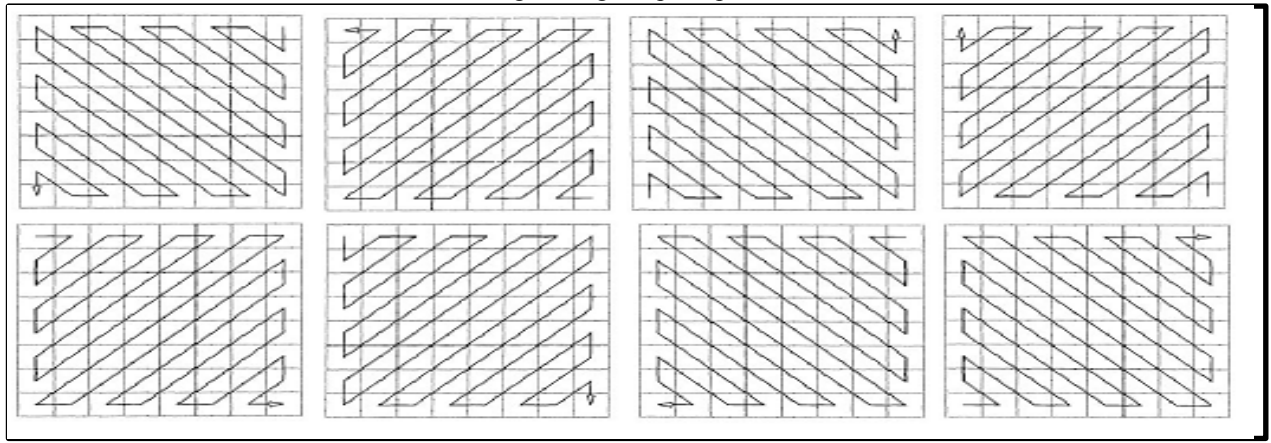

Fig 1. Magic Square pattern

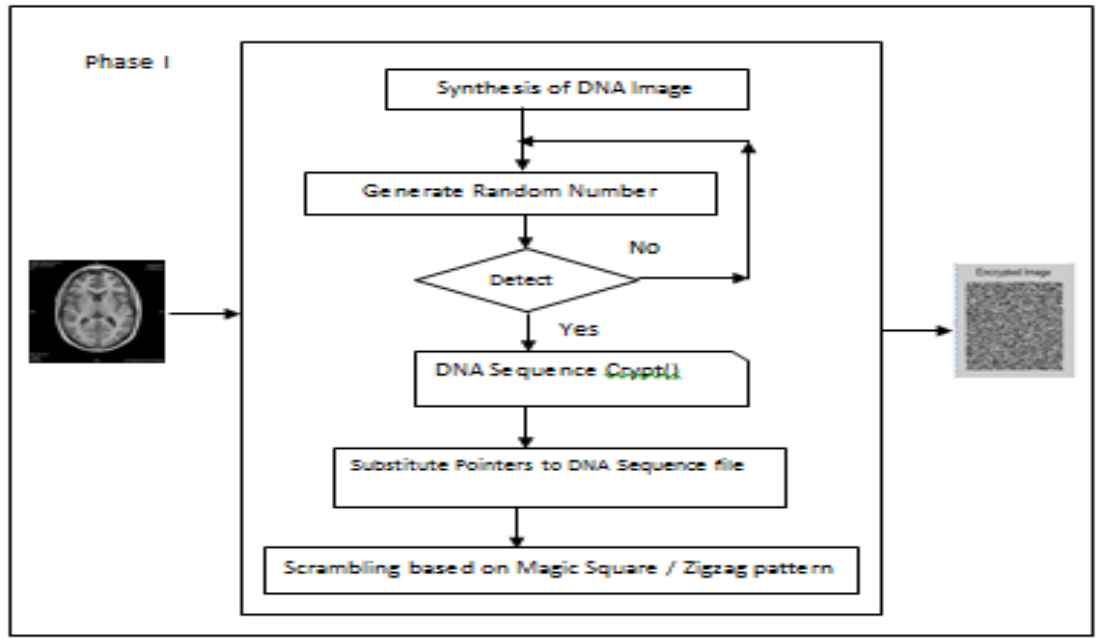

Fig 2. Zigzag pattern

Fig. 3 Phase I

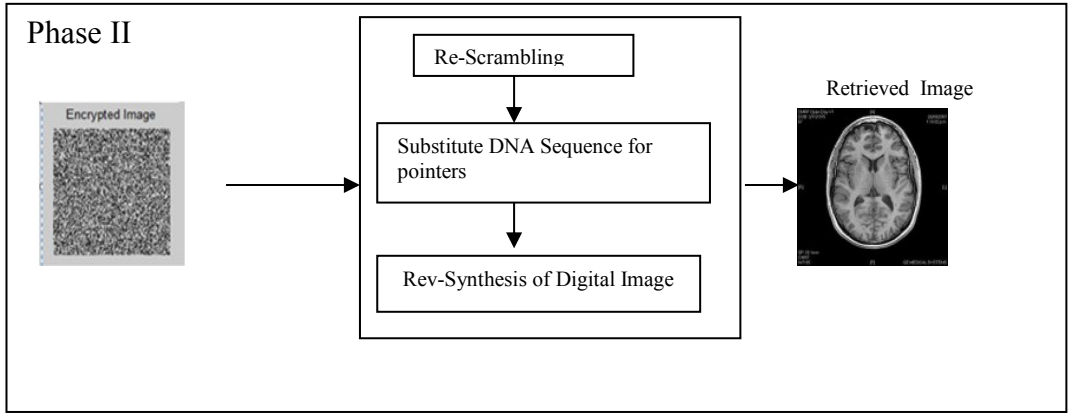

Fig. 4  Phase II

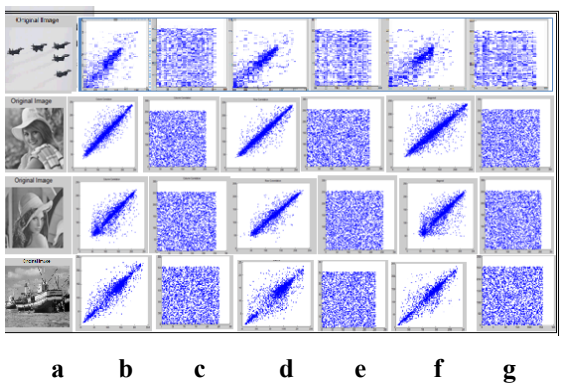

    **a**    **b**    **c**    **d**    **e**    **f**    **g**

Fig. 5 Correlation coefficient of original and encrypted images column-wise, row-wise and diagonal –wise

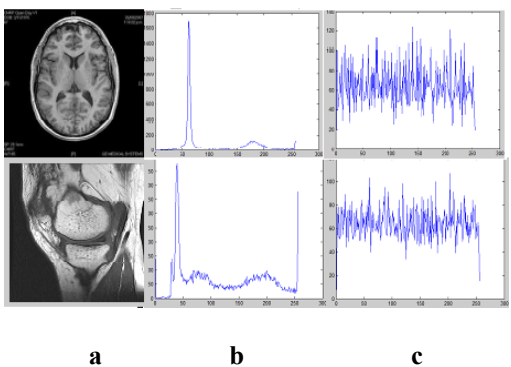

    **a**        **b**        **c**

Fig. 6 brain & bone images – original and encrypted histograms difficult to excerpt the pixels' statistical nature of the original image from the encrypted image and the algorithm can resist a chosen plain image or known plain image attack.

**Table 1. Correlation coefficients**

| Image | Correlation Coefficient | | | | | |
| --- | --- | --- | --- | --- | --- | --- |
| | column-wise | | row-wise | | diagonal-wise | |
| | original | encrypted | original | encrypted | original | encrypted |
| Br.jpg | 0.855 | 0.001 | 0.837 | -0.004 | 0.77 | 0.001 |
| Kn.jpg | 0.86 | 0.006 | 0.959 | -0.01 | 0.854 | 0.011 |

### 4.3. Avalanche Effect

In the integrated encryption scheme, avalanche effect is significantly apparent as a flip in a bit will be mapped to an entirely different position in the key DNA sequence file.

### 4.4 Differential attack

According to Kerckhoffs's principle, the integrated encryption scheme is secure even if everything about the system, except the key, is known to the unauthorized user. As there are a lot of DNA sequences and the scrambling pattern is chosen at random from a look-up table, it is difficult to guess the sequence or the scrambling methodology used ascertaining that the encryption scheme is secure against differential attack.

**4.5 Brute Force Attack**

A Brute Force Attack or exhaustive key search is an approach that involves systematically checking all possible keys until the correct key is found. Applying incorrect DNA digital coding or different DNA sequence or a different scrambling method will cause biological pollution and would lead to a corrupted image.

**5   Conclusion**

The intricacy and unpredictability of DNA based encryption provides a great uncertainty which makes it better than other mechanisms of cryptography. Integrating DNA based encryption along with zigzag ordering or magic square scrambling helps in the double fold secure storage of confidential images. The proposed Integrated Encryption Scheme is easy to implement and can resist statistical, differential and brute-force attack and is suitable for storage of images in multi-cloud.